\documentclass{article}
\pdfoutput=1

\usepackage{amsfonts}
\usepackage{amssymb}
\usepackage{amsmath}
\usepackage{amsthm}
\usepackage{bm}
\usepackage{booktabs}
\usepackage{complexity}
\usepackage{dcolumn}
\usepackage[pdftex]{graphicx,color}
\usepackage{pdfsync}
\usepackage{rotate}
\usepackage{siunitx}
\usepackage{subfigure}
\usepackage{framed}
\usepackage[rgb]{xcolor}
\usepackage{lmodern}
\usepackage[final]{pdfcomment}

\usepackage[pdftex]{hyperref}
\usepackage{authblk}

\newcommand{\avg}[1]{\langle #1 \rangle}
\def\G{\mathcal{G}}

\def\eg{\emph{e.g. }}
\def\ie{\emph{i.e. }}
\def\hop{\mathrm{hop}}
\def\dist{\operatorname{dist}}
\def\naturals{\mathbb{N}}
\newcommand{\bra}[1]{\langle#1|}
\newcommand{\ket}[1]{|#1\rangle}
\newcommand{\braket}[2]{\langle#1|#2\rangle}
\newcommand{\iprod}[3]{\left\langle #1 \rvert #2 \lvert #3 \right\rangle}
\begin{document}

\title{Ergodic and localized regions in quantum spin glasses on the Bethe lattice}

\author[1,2]{G. Mossi}
\author[2,3]{A. Scardicchio}

\affil[1]{SISSA, via Bonomea 265, 34136 Trieste, Italy}
\affil[2]{INFN, Sezione di Trieste, Via Valerio 2, 34127 Trieste, Italy}
\affil[3]{Abdus Salam ICTP, Strada Costiera 11, 34151 Trieste, Italy}

\date{}
\renewcommand\Affilfont{\itshape\small}

\maketitle

\begin{abstract}
By considering the quantum dynamics of a transverse field Ising spin glass on the Bethe lattice we find the existence of a many body localized region at small transverse field and low temperature. The region is located within the thermodynamic spin glass phase. Accordingly, we conjecture that quantum dynamics inside the glassy region is split in a small MBL and a large delocalized (but not necessarily ergodic) region. This has implications for the analysis of the performance of quantum adiabatic algorithms.
\end{abstract}


\section{Introduction}
\setstcolor{blue}

In recent years the study of the different dynamical regimes of isolated quantum systems has received a lot of attention, due to  improved experimental techniques \cite{ColdAtoms, SCQB} and theoretical progress. In the latter, one can identify two distinguished but not independent lines of research: the first is the study of how ergodicity is realized in isolated quantum systems, a mechanism that goes under the name of \textit{eigenstate thermalization hypothesis} \cite{ETH}; The second is the study of the most typical mechanism for failure of ergodicity in presence of quenched disorder \cite{A58,BAA,PH,HNO1,ZPP} (although some authors have suggested disorder in the initial state suffices \cite{SM,DU1}) named many-body-localization. While dynamical phases satisfying ETH can be described by the usual tools of statistical mechanics and thermodynamics, MBL systems behave a lot like integrable systems \cite{BDS} with local integrals of motions \cite{HNO1,SPA,I16,RMS,IRS}: Transport is suppressed \cite{DS,LLA}, entanglement entropy grows slowly to its thermodynamic value \cite{BPM}, and some symmetry breaking phases can exist, protected from the Mermin-Wagner theorem, in low dimensions and at high temperature \cite{CKLS}.

Soon after its inception, it has been pointed out that MBL phases can be detrimental \cite{AKR} for the performance of Adiabatic Quantum Computation (AQC) protocol introduced in \cite{FGG} (see also \cite{KN}). This has been contested in later works \cite{KS} and it remains a controversial claim. Since this protocol has proven to be the most promising for the realization of a quantum computer \cite{LMSS}, sorting out this question is of paramount importance for both theoretical discussions and technological implications.

In a series of recent works, which involve one of the present authors \cite{LPS, BLPS1, BLPS2}, the question of the appearance of an MBL phase in some models of quantum spin glasses has been addressed with the result that, for realistic, mean-field glasses MBL can exist only in finite-connectivity models, while in fully-connected models only a weaker form, remnant of the clustering phase existing in the phase space of the classical model \cite{MPV}, exists. These earlier works point at the necessity to examine a finite-connectivity quantum spin glass, in search of MBL.

In this work we set to do exactly this: we analyze the quantum dynamics of an isolated quantum spin glass showing that there is an ETH phase at large transverse field, possibly extending all the way down to the spin-glass phase while at all system sizes we were able to study, an MBL phase exists for small transverse field. Therefore there should be an MBL-ETH dynamical transition in between. 

\section{Thermodynamics}

The focus of the present paper is the transverse-field Ising spin glass model defined on a regular random graph (RRG) of degree $d=3$, whose Hamiltonian is defined by
\begin{equation}\label{eq:hamiltonian}
H = -\sum_{\avg{i,j}} J_{ij} \sigma_i^z \sigma_j^z - \Gamma \sum_{i=1}^N \sigma_i^x,
\end{equation}
where $\Gamma > 0$ is the strength of the transverse field, the disordered interaction couplings $J_{ij}$ take either of the two values in $\{\pm 1\}$ with equal probability and $\sigma_i^a$ (for $a = x,z$) is a Pauli matrix acting on the $i$-th spin of the system. The sum of the terms $\avg{i,j}$ is taken over the edges of a $3$-regular reandom graph $\G$. Consequently, the Hamiltonian of Eq. \ref{eq:hamiltonian} has two sources of disorder: the disordered interactions $J_{ij}$ and the topology of the underlying regular random graph $\G$.

\begin{figure}[htp]

\centering
\includegraphics[width=.3\textwidth]{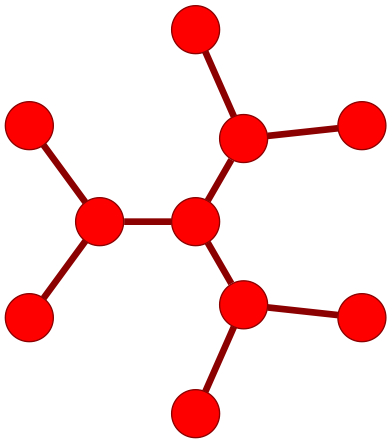}\hfill
\includegraphics[width=.3\textwidth]{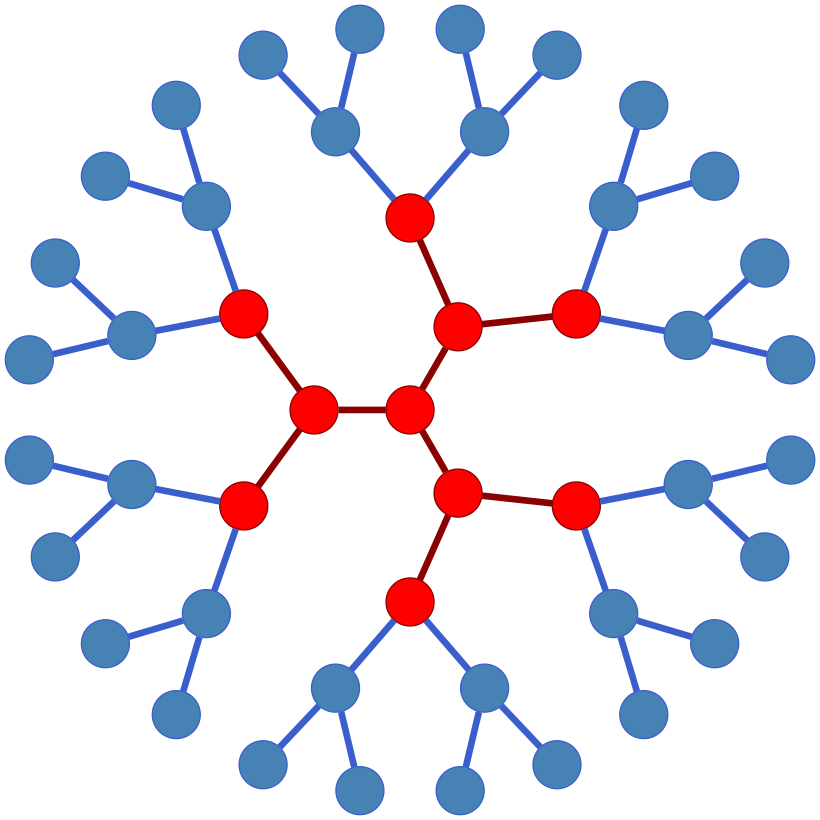}\hfill
\includegraphics[width=.3\textwidth]{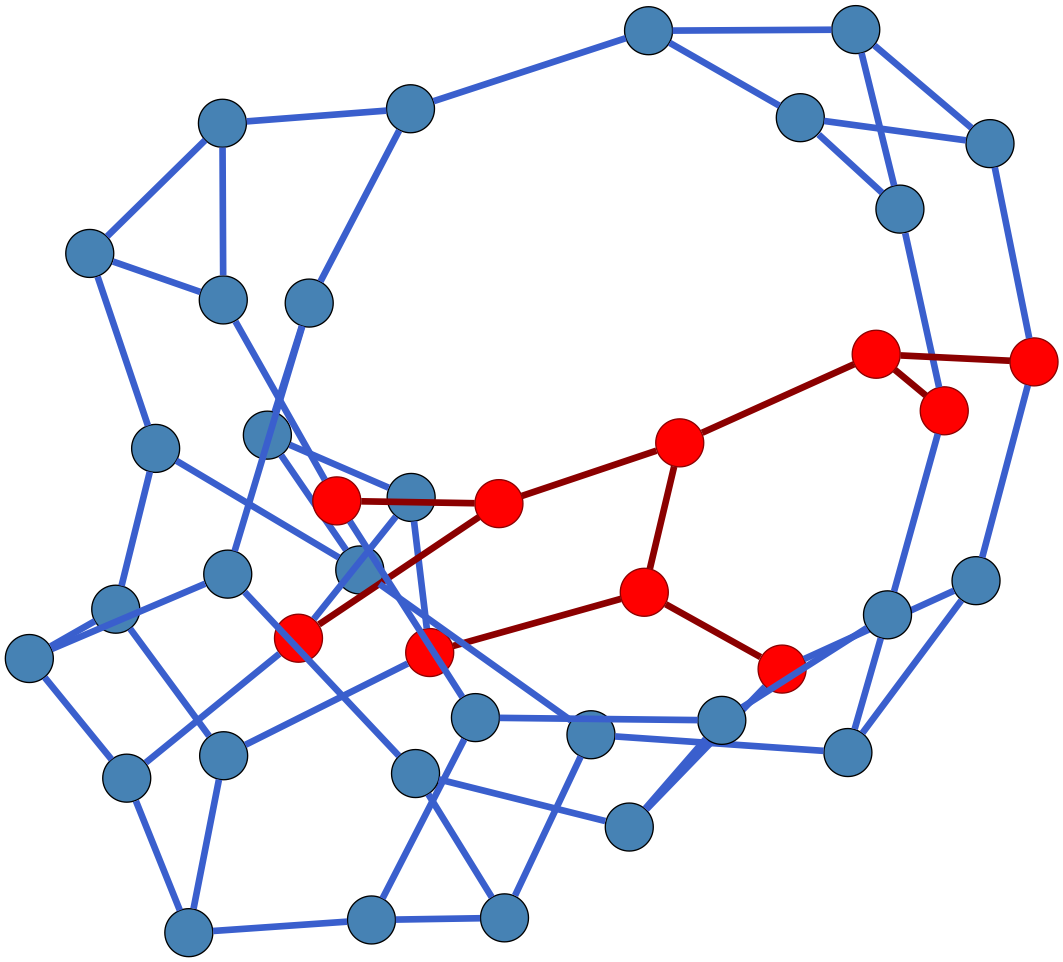}

\caption{Locally, both the Bethe lattice and a regular random graph have a tree structure. The loops that are present in a regular random graphs appear only at the global scale.}
\label{fig:locally_tree_like}

\end{figure}

\noindent Here and in what follows, a $d$-regular\footnote{the degree of a regular graph is also customarily defined by referring to its ``connectivity'' $K$. The relations between degree and connectivity is given by $K = d-1$.} random graph of size $N$ is defined as graph uniformly sampled from the set $\mathbb{G}_N^d$ of all connected graphs with $N$ vertices and fixed degree $d$. These graphs are related to the $d$-regular Bethe lattice, the unique (up to isomorphism) connected tree graph of fixed degree $d$ with denumerably-many vertices. The Bethe lattice was introduced in \cite{Bethe1935} to define models where the Bethe-Peierls approximation is exact. Regular graphs can be seen as finite-size approximation to the Bethe lattice in the sense that even though they contain loops (and the Bethe lattice does not) the fraction of loops of any fixed length $\ell$ vanishes as the size $N$ approaches infinity (\ie almost all loops are of length greater than $\ell$). Thus RRGs are ``tree-like'' in the (finite-sized) neighbourhood of any of their vertices, and are locally indistinguishable from the Bethe lattice (see Fig. \ref{fig:locally_tree_like}). For this reason, regular random graphs in the $N \rightarrow \infty$ limit have been used to model the statistical properties of the Bethe lattice and approximation schemes related to the Bethe-Peierls method (such as the cavity method and belief-propagation algorithms) are expected to perform well when applied to systems with local interactions that define a regular random graph structure.

\begin{figure}
\begin{center}
  \includegraphics[width=.49\linewidth]{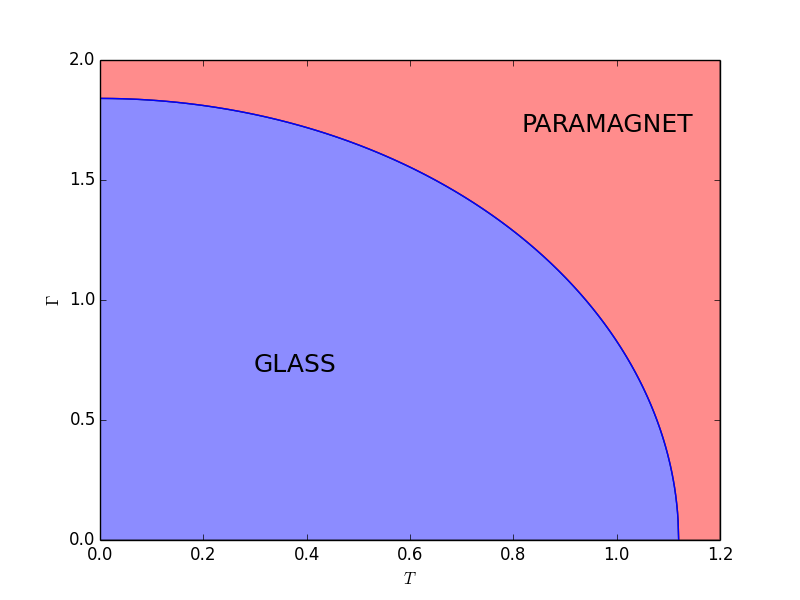}
\caption{Phase diagram of the transverse-field Ising spin glass model of Eq. \ref{eq:hamiltonian}. The model transitions from
a paramagnetic to a glassy phase for small values of the transverse field $\Gamma$ and temperature $T$.}
  \label{fig:phase_diagram}
\end{center}
\end{figure}

The thermodynamical properties of the Hamiltonian of Eq. \eqref{eq:hamiltonian} were studied in a series of papers \cite{Laumann2008,Farhi2012} that collectively reconstructed the equilibrium phase diagram shown in Fig. \ref{fig:phase_diagram}. For large values of the transverse field, or for high temperatures, the system lies in a paramagnetic phase where the $z$-component of each spin fluctuates randomly, so that the average $z$-magnetization is zero. In the region of small values of $\Gamma$ and low temperatures $T$, each spin freezes independently in the $z$-direction and the system enters a phase where the Edwards-Anderson order parameter,
\[
q_{EA} \equiv \frac{1}{N}\sum_{i=1}^N \avg{\sigma_i^z}^2,
\]
becomes strictly positive with a continuous transition \cite{Laumann2008,Mossi2016}.
\noindent The classical $\Gamma = 0$ line of the phase diagram, where the transition is driven solely by the temperature, was studied in \cite{VianaBray85} where it was found that the critical temperature was given by the formula $T_c = 1/\tanh^{-1}(1/\sqrt{2}) \approx 1.13$. The interior of the $(T,\Gamma)$-plane was studied in Ref. \cite{Laumann2008}. The authors developed a quantum version of the cavity method to explore numerically the thermodynamic limit of the model and compute the critical line. The physics of the quantum $T = 0$ line was studied by a series of papers \cite{Laumann2008,Farhi2012} that gave various estimates of the critical point using cavity-like approximation schemes affected by the systematic error that these methods exhibit when applied to loopy graphs. It was conclusively studied numerically in \cite{Mossi2016} where the critical value of the transverse field was computed by path-integral Monte Carlo methods using several different physical quantities. All were found to agree on an estimated value of $\Gamma_c = 1.82 \pm 0.02$ with a finite-size correction that disappears as $1/N$ for $N \rightarrow \infty$.  Zero-temperature properties of the ground state were also studied in \cite{Mossi2016} where the ground state was found to have volumetric entanglement -- as measured by the R\'enyi entropy of order two -- in the entire part of the $T=0$ line which lies in the paramagnetic phase. The equal-time spin-spin connected correlation function in the $z$ direction,
\[
C_{ij} \equiv \avg{\sigma_i^z\sigma_j^z} - \avg{\sigma_i^z}\avg{\sigma_j^z},
\]
was also studied in \cite{Mossi2016} in order to analyze the average and the extremal behaviour of the correlations across the transition. They considered the spatial maximal correlation $C_{\max}(r)$, defined by picking the maximal value of $C_{ij}$ among all spins $j$ that are at fixed distance $r$ from a given spin $i$ (and averaging over $i$), and the mean correlation $C_{\mathrm{mean}}(r)$, defined by taking the average value of $C_{ij}$ among all spins $j$ that are at fixed distance $r$ from $i$. Both were shown to follow a stretched exponential decay, and the correlation lengths extracted from them were found to converge to a finite value at the critical point of the transition in the thermodynamic limit.

\section{Ergodic region at large transverse field}

In order to develop some intuition about the phase transition, the authors of Ref. \cite{Mossi2016} did a perturbative expansion by using the modulus of the interaction strength $J = |J_{ij}|$ as the perturbative parameter. For ease of reference we summarize here their results. The ``free'' Hamiltonian is the transverse-field term $H_0 \equiv -\Gamma\sum_i \sigma_i^x$. By shifting the ground state energy $E_0=-\Gamma N$ to zero, its spectrum is given by
\[
\operatorname{Sp}(H_0) = \{0,2\Gamma, 4\Gamma, \ldots, 2N\Gamma\}.
\]
Its unique ground state $\ket{0}$ (where we denote $\sigma^x\ket{\rightarrow} = \ket{\rightarrow}$)
\[
\ket{0} \equiv \bigotimes_{i=1}^N \ket{\rightarrow}_i,
\]
is taken as a pseudovacuum of quasiparticles. The operators $\sigma_i^z$ create an excitation on top of the ground state
\[
\ket{i} \equiv \sigma_i^z \ket{0} = \ket{\leftarrow}_i \otimes \Big( \bigotimes_{j \neq i}\ket{\rightarrow}_j \Big),
\]
which is interpreted as a state containing a quasiparticle at site $i$. These form the $N$-fold degenerate eigenspace of $H_0$ with energy $2\Gamma$. Additional applications of the $\{ \sigma_j^z \}$ operators either move the state upwards in the spectrum, creating states with two, three, or more quasiparticles $\ket{i,j},\ket{i,j,k},\ldots$, or downwards, by annihilating existing quasiparticles.

\noindent The perturbed Hamiltonian is
\[
H(J) = H_0 + JV
\]
where $V$ is the dimensionless spin-glass term $V = -(1/J)\sum_{\avg{i,j}} J_{ij} \sigma_i^z \sigma_j^z$

\noindent First-order perturbation theory gives a null correction to the ground-state energy
\[
E_0(J) = J E_0^{(1)} = 0
\]
since $JE_0^{(1)} = \avg{0 \rvert JV \lvert 0} = 0$. The degenerate band of one-particle states is split by the perturbation into distinct levels
\begin{equation}\label{eq:first_order_energy_correction}
E_n(J) = 2\Gamma + J E_n^{(1)}
\end{equation}
for $n = 1, \ldots, N$, where $J E_n$ are eigenvalues and $\{\ket{\phi_n}\}$ are the eigenvectors of the operator $\tilde{V}$, which is the perturbation operator $V$ restricted to the (unperturbed) one-particle subspace. The direct computation of the $E_n^{(1)}$ requires the diagonalization of an $N$ by $N$ matrix which, we will show, is a hopping matrix on the RRG.

\noindent Let us start by noticing that the term $\sigma_i^z \sigma_j^z$ applied to a quasiparticle state $\ket{i',j',k',\ldots}$ can affect it in one of three ways: (i) it can create a pair of adjacent quasiparticles provided the sites $i$ and $j$ are devoid of them, (ii) it can move a quasiparticle from site $i$ to site $j$ provided site $j$ is empty and site $i$ is occupied (or vice versa, swapping the role of $i$ and $j$), or (iii) it can annihilate a pair of adjacent quasiparticles sitting on sites $i$ and $j$.

\begin{figure}[htp]

\centering
\includegraphics[width=.45\textwidth]{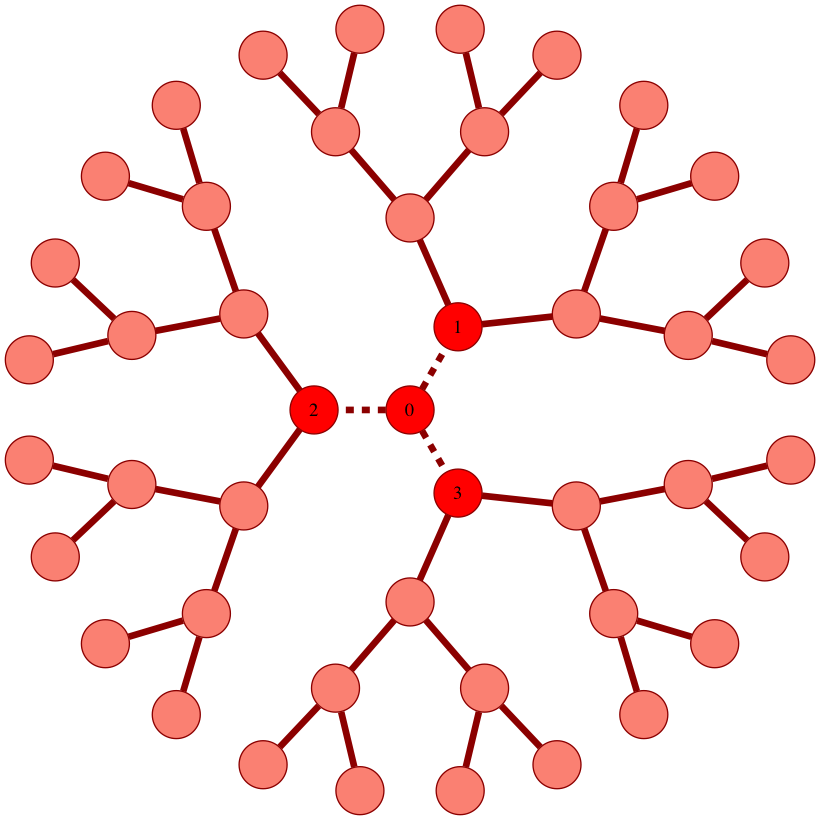}\hfill
\includegraphics[width=.48\textwidth]{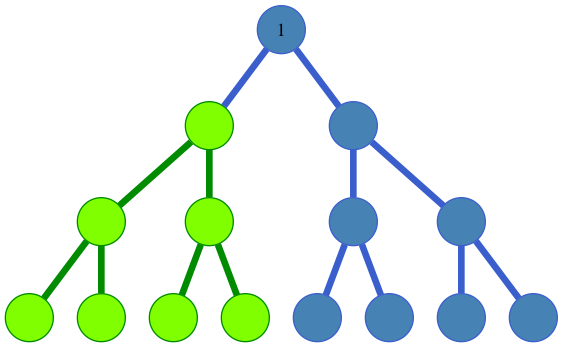}

\caption{(Left) The diagonal Green's function $G_i$ at the point $i=0$ on a $3$-regular Bethe lattice can be computed once the values $G^{(c)}_j$ are known at the points $j = 1,2,3$ (shown in red). $G_j$ is the Green's function, computed at the site $j$, of the operator obtained from the Hamiltonian $H_{\hop}$ by removing the hopping terms terms associated to the dashed edges of the graph. (Right) Each disconnected component thus obtained is an infinite binary tree that is isomorphic to all its infinite descending binary subtrees (one such subtree is shown in green).}
  \label{fig:cavity}
\end{figure}

\noindent Note also that in the one-particle subspace annihilation processes cannot happen, while creation processes map a state into the three-particle subspace, which is orthogonal to the one-particle subspace. Therefore, only the hopping processes give a contribution to Eq. \eqref{eq:first_order_energy_correction}. For any state $\ket{\psi}$ in the one-particle subspace, the action of  $J\tilde{V}$ is then equivalent to that of $H_{\hop}$, the Hamiltonian of a particle hopping on the same graph $\G$, with disordered hopping constants $J_{ij}=\pm J$
\[
H_{\hop} = -\sum_{\avg{i,j}} J_{ij} \Big( \ket{i}\bra{j} + \ket{j}\bra{i} \Big).
\]
For contrast, consider the Hamiltonian $H_{\hop}^{(\mathrm{hom})} = - J \sum_{\avg{i,j}} \Big(\ket{i}\bra{j} +\ket{j}\bra{i} \Big)$ of a particle hopping on the graph $\G$ with homogeneous hopping coefficients $J>0$. Even though solving for the spectrum of the latter Hamiltonian is difficult for any finite graph $\G$, in the thermodynamic limit, our RRG $\G$ is the Bethe lattice and so spectral properties of this model can be computed exactly using an iterative method (see \eg \cite{A-CAT1,Aizenman2013} and Fig. \ref{fig:cavity}). In particular, its spectral density is known to be supported on the set $[-2J\sqrt{K},2J\sqrt{K}]$, where $K$ is the constant connectivity of the graph. We give the proof here: one starts by writing down iteration equations for the diagonal Green's function $G_i \equiv \iprod{i}{(E-H_{\hop})^{-1}}{i}$ at the site $i$ for generic complex $E$
\begin{equation}
\label{eq:iterG}
G_i=\left(E-\sum_{j\in\partial i}J_{ij}^2 G^{(c)}_j\right)^{-1},
\end{equation}
where the cavity Green's function $G^{(c)}_j = \iprod{j}{(E-H^{(i)}_{\hop})^{-1}}{j}$ is the Green's function of the operator $H^{(i)}_{\hop}$ obtained from the Hamiltonian $H_{\hop}$ by removing the hopping terms associated to the the edges incident to the site $i$. On the Bethe lattice the removal of these terms splits the system into $K+1$ isomorphic disconnected components that can be considered independently. Each component is an infinite rooted tree with a branching factor of $K$. These trees are isomorphic to each of their infinite descending subtrees rooted at any of their vertices. By writing the iteration equation \eqref{eq:iterG} for $G_j^{(c)}$ one gets
\begin{equation}
\label{eq:tree_propagator}
G^{(c)}_j=\left(E-\sum_{k \in \partial j}J_{jk}^2 G^{(c')}_k\right)^{-1},
\end{equation}
where $G^{(c')}_k$ is a second-step cavity Green's functions obtained from the $G^{(c)}_j$ Hamiltonian by further removing the hopping terms associated to the edges incident to $j$. By solving \eqref{eq:tree_propagator} and plugging the result in \eqref{eq:iterG} one recovers $G_i$ and from $G_i$ the spectral density
\[
\rho_i(E) = \frac{1}{\pi} \lim_{\Im E \rightarrow 0^+} \Im G_i(E).
\]

\noindent One can find $P(\Im \Sigma_i)$ by $\Sigma_i(E)=E-G_i^{-1}$ and taking the limit $\Im{E}\to 0$. As we know from \cite{A58}, the distribution of $\Im \Sigma$ will tend to a delta function for \emph{delocalized states} and to a long-tailed distribution for \emph{localized states}. However we can sidestep all this procedure recognizing that in all the equations only $J_{ij}^2$ appears and therefore the case $J_{ij}=\pm J$ is \emph{identical} to the case $J_{ij}=J$ constant, which has only delocalized states. The latter statement follows from the observation that, given the constant connectivity of the graph and the constant value of the hopping $J$ the distribution of $G^{(c)}$ must be a delta function centered on the solution of the deterministic equation (\ref{eq:iterG})
\begin{equation}
G^{(c)}=\left(E-KJ^2 G^{(c)}\right)^{-1},
\end{equation}
which gives
\begin{equation}\label{eq:cavity_solution}
G^{(c)}=\frac{2}{E-\sqrt{E^2-4KJ^2}}.
\end{equation}
By inserting \eqref{eq:cavity_solution} into \eqref{eq:iterG} one gets
\begin{eqnarray*}
	G_i^{-1} &=& E- (K+1) J^2 G^{(c)}\\ &=& E - \frac{2(K+1)J^2}{E-\sqrt{E^2-4KJ^2}}
\end{eqnarray*}
so
\[
\Sigma_i = \frac{2(K+1)J^2}{E-\sqrt{E^2-4KJ^2}}
\]
irrespective of $i$ and therefore all the eigenstates are \emph{delocalized}.

By extending this reasoning to any sector with $n$ particles, as soon as $n$ is finite when $N\to\infty$ we can conclude that all such few-particle states are delocalized. As $n/N$ becomes appreciable so does the interaction between the particle, we shall conjecture that the introduction of a small interaction between the particles (which is of $O(J^2/\Gamma)$), irrespective of its attractive or repulsive nature, does not \emph{localize} excitations (this is true even in the presence of bound states) and the phase remains delocalized. The value of $\Gamma/J$ where this breaks down we cannot predict without considering quantitatively the interaction between the particles.

Having showed that there is an ergodic region for large $\Gamma/J$ we will show in the following section that starting from the opposite limit $\Gamma/J\ll 1$ we do have a localized region. Therefore there must be at least one dynamical phase transition in between. We unfortunately are not able to answer the question whether this transition is unique or a crossover through a sequence of ergodicity breaking transitions. 

\section{Many-body localization at small transverse field}

\subsection{Localization and the Forward Approximation}

Let us consider the Hamiltonian \eqref{eq:hamiltonian} for small values of $\Gamma$. Then $H = H(\Gamma)$ can be treated as a small perturbation of the spin glass term $H_0 \equiv -\sum_{\avg{i,j}}J_{ij}\sigma_i^z \sigma_j^z$ with the transverse-field operator $V = -\sum_i \sigma_i^x$ acting as the perturbation:
\[
H(\Gamma) = H_0 + \Gamma V.
\]
Note that $H_0 = -\sum_{\avg{i,j}} J_{ij}\sigma_i^z \sigma_j^z$ is diagonal in the $N$-fold $\sigma^z$ product basis, whose elements we label by strings $a \in \{\pm 1\}^N$ through the usual identification $\ket{\uparrow} \equiv \ket{1}$ and $\ket{\downarrow} \equiv \ket{-1}$.
Now and throughout this section we use Latin letters $a,b,c,\ldots, x,y,z$ to label energies and eigenstates of the unperturbed Hamiltonian $H_0$ while Greek letters $\alpha,\beta,\gamma,\ldots$ are reserved as labels for the energies and the eigenstates of the perturbed Hamiltonian, so that in the $\Gamma\rightarrow 0$ limit the perturbed Greek labels ``converge'' to their corresponding Latin label\footnote{while this convergence does not hold in general for any choice of eigenbasis for the degenerate Hamiltonian $H_0$ we will later show that in our case this is the correct choice.}, \eg for the energies $E_{\alpha} \rightarrow E_a$ and for the energy eigenstates $\ket{\psi_\alpha} \rightarrow \ket{a}$.

\noindent The eigenstates $\ket{\psi(\Gamma)}$ of the Hamiltonian $H(\Gamma)$ can be chosen to depend continuously on the parameter $\Gamma$, and will converge to the eigenstates of $H_0$ in the limit $\Gamma \rightarrow 0$. Consequently, if we write the wavefunctions of the states $\ket{\psi(\Gamma)}$ in the unpertubed energy eigenbasis
\[
\psi(x) = \braket{x}{\psi(\Gamma)}
\]
we expect that these will converge to Kronecker delta functions as $\ket{\psi(\Gamma)}$ changes into the corresponding unperturbed eigenstate $\ket{a}$; this means that the state $\ket{\psi(\Gamma)}$ will be localized for small $\Gamma$. In the previous section we saw that in the large $\Gamma$ limit these states are delocalized so we expect that there will be a value $\Gamma_c$ where transition between these two behaviours occours. In fact, if we write the Hamiltonian is the eigenbasis of $H_0$ we get
\begin{equation}\label{eq:altshuler_model}
H = -\sum_{a} E_{a}\ket{a}\bra{a} - \Gamma \sum_{a,a'} \iprod{a}{V}{a'} \ket{a}\bra{a'}.
\end{equation}
This is related to the Anderson model that describes a particle hopping on a lattice while under the effect of a disordered potential. This model is known to enter a localized phase when the disordered term dominates the hopping term. In Eq. \eqref{eq:altshuler_model} the unperturbed energies $\{E_a\}$ are the equivalent of a disordered potential while the values $\Gamma\iprod{a}{V}{a'}$ are the hopping coefficients for the particle. Note that here the underlying geometry of the hopping is defined by the matrix elements $\iprod{a}{V}{a'}$: the sites $a,a'$ are adjacent (\ie the particle can hop directly from one to the other) if and only if $\iprod{a}{V}{a'} \neq 0$. As $\Gamma \rightarrow 0$ the hopping is suppressed and the system is likely to enter a disorder-induced localized phase. In the transverse-field Ising spin glass Hamiltonian \eqref{eq:hamiltonian} the perturbation $V = -\sum_i \sigma_i^x$ defines a hopping over the $N$-dimensional Boolean hypercube $\mathcal{B}_N = \{\pm 1\}^N$.

A disorder-driven localization/delocalization transition is usually studied as a function of two quantities: the energy density $\epsilon$ of the eigenstates considered and the strength $W$ of the disorder in the Hamiltonian. One usually finds that states at a fixed energy density $\epsilon$ change from delocalized to localized as the disorder strength $W$ is increased, with an energy-dependent critial value $W_c(\epsilon)$ that marks the boundary between these two behaviours. The set of critical points $W_c(\epsilon)$ for different values of $\epsilon$ define the ``mobility edge'' of the system.

In our setup we keep the strength of the disordered interactions fixed $W \equiv \lvert J_{ij}\rvert = 1$ so the relative strength of the disorder with respect to the ordering term $-\Gamma\sum_i \sigma_i^x$ is controlled by the parameter $W/\Gamma = 1/\Gamma$. The mobility edge will consequently be defined by the critical values $\Gamma_c(\epsilon) \equiv 1/W_c(\epsilon)$.

Let us better define the kind of localization we will study. First, we fix a value $\epsilon$ for the energy density and $\Gamma$ for the strength of the transverse field. For a given realization of disorder of the Hamiltonian $H(\Gamma)$ \eqref{eq:hamiltonian} defined on a system of size $N$ and a state $\ket{\psi}$ with localization center $\ket{v}$ and energy density $\epsilon$ we define
\begin{equation}\label{eq:psi_r}
\psi_r \equiv \max\Big\{\lvert \braket{w}{\psi} \rvert : w \in \{\pm 1\}^N, \dist(v,w) = r\Big\} \quad \quad r = 1,2,\ldots
\end{equation}
where the distance is taken over the Boolean hypercube. Note that in order to compute $\psi_r$ from the amplitudes $\lvert \braket{w}{\psi} \rvert$ one needs to know (or find out) the localization center of the state $\ket{\psi}$. Next, for fixed $N,r \in \naturals$ and real numbers $C,\xi > 0$ we define the quantity $P(N,r,\xi)$ as the probability (taken over all disorder realizations of $H(\Gamma)$ of size $N$ and states of energy density $\epsilon$) that the random variable $\lvert \psi_r\rvert^2$ satisfies $\lvert \psi_r\rvert^2 \leq Ce^{-r/\xi}$
\[
P(N,r,\xi) \equiv Pr \Big( \lvert \psi_r \rvert^2 \leq Ce^{-r/\xi} \Big).
\]
We say that a disordered system is localized if there exist a real number $\xi>0$ such that
\begin{eqnarray*}
 \lim_{r \rightarrow \infty} \lim_{N \rightarrow \infty} P(N,r,\xi) = 1.
\end{eqnarray*}
This means that the probability distribution $p(x) = \lvert \psi(x) \lvert^2$ is under an exponentially-decaying envelope, except possibly for a region of finite radius.

\noindent Notice that we can write equivalently
\[
 P(N,r,\xi) = P\Big( \frac{\ln\lvert\psi_r\lvert^2}{2r} \leq -\frac{1}{2\xi} + O\Big(\frac{1}{r}\Big)\Big),
\]
so we can study the distribution of the values of the random variable $Z_r \equiv \ln\lvert\psi_r\lvert/r$. It was observed \cite{Pietracaprina2016} that if the system is localized then in the limit $N \rightarrow \infty$ the random variable $Z_N$ is peaked around the value
\begin{equation}\label{eq:gamma_c}
Z_N \equiv \frac{\ln\lvert\psi_N\lvert}{N} \rightarrow  \ln (\Gamma/\Gamma_c) \quad \quad \text{as $N \rightarrow \infty$},
\end{equation}
which gives the value of $\Gamma_c(\epsilon)$. In practice one can calculate $\psi$ with $\Gamma=1$ and get $\Gamma_c$. By the way the relation $\xi=1/\ln(\Gamma/\Gamma_c)$ tells us the critical exponent for the $\xi$ divergence is 1, since $\xi\sim1/|\Gamma-\Gamma_c|$.

We have seen how one can study the localization properties of disordered systems by looking at the wavefunction values $\psi(x)$. Of course this is not always a simple task, as it usually requires the diagonalization of a matrix whose size grows exponentially with the size of the system. However, in a perturbative setup such as the one we described we can use a technique known as the ``forward approximation'' \cite{A58,Pietracaprina2016}, namely, we neglect the renormalization of the free energy of the unperturbed eigenstates $\{\ket{a}\}$. This is known to give an \emph{underestimate} of the critical hopping strength both in the many-body and Anderson localization, it is then useful for our purpose of proving the stability of the phase (basically the same strategy is used in \cite{BAA,RMS}).

The forward approximation states that for a perturbed energy eigenstate  $\ket{\psi_\alpha}$ that converges to $\ket{a}$ at $\Gamma = 0$, the value of the wavefunction $\psi_\alpha(b) = \braket{b}{\psi_\alpha}$ can be approximated by a sum of contributions associated to the shortest paths connecting the sites $a$ and $b$ in the Boolean hypercube:
\begin{equation}
\label{eq:forwardWF}
\psi_\alpha(b) \approx \sum_{p\in\text{spaths}(a,b)} \, \prod_{i \in p} \, \frac{\Gamma}{E_a-E_i},
\end{equation}
where the set $\text{spaths}(a,b)$ contains the shortest paths from $a$ to $b$.


Notice that the sum over paths of Eq. \eqref{eq:forwardWF} can be computed numerically using a transfer matrix technique, in which case one computes iteratively the vector
\begin{equation}
v_t=D\cdot A\cdot v_{t-1}
\end{equation}
with
\begin{eqnarray}
D_{bc}&=&\delta_{bc}\frac{1}{E_a-E_b},\\
A_{bc}&=&(V)_{bc},\\
(v_0)_c&=&\delta_{ac}
\end{eqnarray}
and
\begin{equation}
\psi_\alpha(b)=(v_N)_b,
\end{equation}
where $N$ is the distance between $a$ and $b$ in the Boolean hypercube. One can decrease memory requirements by noting that the vector $v$ is very sparse during most of the computation, so at each step most entries of the transfer matrix $\mathcal{T} = DA$ are irrelevant. This is because repeated applications of $\mathcal{T}$ to the initial state define a diffusion process on the Boolean hypercube where at each step $t$ one needs to propagate only the amplitudes of the vertices exactly at distance $t$ from the initial vertex. This means that in practice one does not need to store in memory the entire transfer matrix $\mathcal{T}$, but instead a new transfer matrix $\mathcal{T}_t$ is defined at each step that only propagates amplitudes from vertices actually relevant for that single step of propagation. This requires storing only $(N-t)\binom{N}{t}$ non-zero entries instead of $N2^{N}$ of the full transfer matrix $\mathcal{T} = DA$. The vector $v_t$ need to store only $\binom{N}{t}$ entries.

\subsection{Numerical results}
In this section we apply the previously-described methods to the transverse-field Ising spin glass Hamiltonian \eqref{eq:hamiltonian}. Here we immediately face an issue: the computation of the forward approximation is obstructed by the fact that the spin glass term $H_0 = -\sum_i J_{ij} \sigma_i^z \sigma_j^z$ has highly degenerate energy levels. This gives rise to diverging terms in Eq. \eqref{eq:forwardWF} when $E_a = E_i$. In order to avoid this problem we add a weak and random longitudinal field term $H_{\mathrm{long}}$ to the Hamiltonian $H_0$:
\[
H_{\mathrm{long}} = - \sum_i h_i \sigma^z_i
\]
where each $h_i$ is distributed uniformly in $(-h, h)$ with $h = 0.001$ (this has to be $\ll 1/N$ but $\gg e^{-a N}$). This has the effect of splitting the degeneracies while introducing only a negligible effect in the energies of the configurations (and therefore in the amplitudes $\psi_\alpha(b)$) and on the amplitudes of transition between non degenerate states.

We compute the many-body mobility edge for the system in the following way. For each system size $N = 18, 20, 22, 24, 26$ we randomly generate a suitable number of realizations of disorder and for each of these we generate a set of initial states $a = a_1,a_2,\ldots,a_k$, making sure that their energy densities $\epsilon_a = E_a/N$ are (approximately) uniformly distributed in the range allowed by the model. Each of these states $a$ is then propagated to its $\mathbb{Z}_2$-symmetric state $b = -a$ (global spin flip) using the forward approximation algorithm with fixed $\Gamma = 1$ in order to compute
\[
\psi_\alpha (-a)
\]
Notice that the configuration $-a$ is the only configuration $b \in \{\pm 1\}^N$ to satisfy $\dist(a,b) = N$, ergo for the state $\ket{\psi_\alpha}$ we read from Eq. \eqref{eq:psi_r} that (after setting $r=N$)
\[
\psi_N = \max \Big\{ \braket{b}{\psi_\alpha} : b \in \{\pm 1\}^N, \dist(a,b) = N \Big\} = \lvert \psi_\alpha(-a) \rvert.
\]
The results are then binned according to the energy density of the initial state $a$ and the average of the random variable $Z_N = \ln \lvert \psi_N\lvert /N$ value was taken for each bin, obtaining $Z_N(\epsilon) = \avg{Z_N}$.

\begin{figure}[htbp]
\begin{center}
\includegraphics[width=0.9\columnwidth]{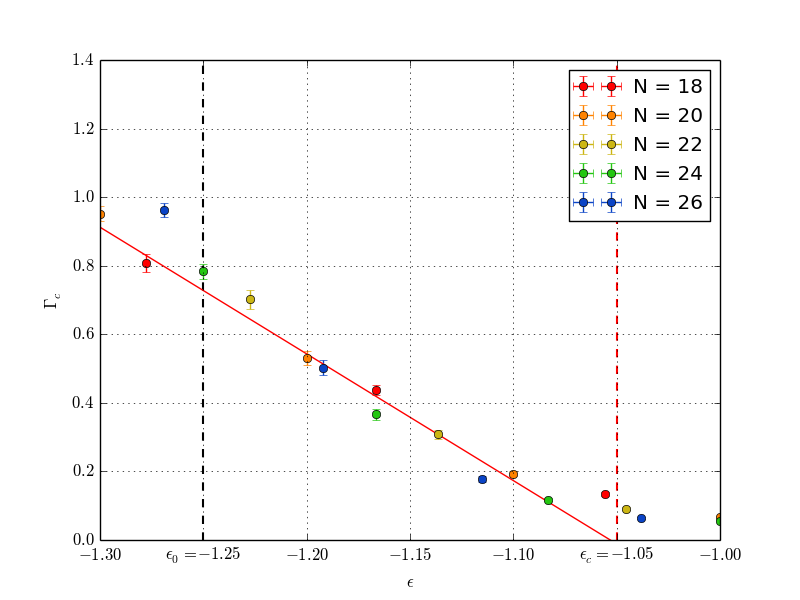}
\caption{Finite-size estimates for $\Gamma_c$ as a function of the unperturbed energy density $\epsilon$, obtained using the forward approximation together with a linear fit of the data at largest $N$. The ground-state energy density of the unperturbed model in the thermodynamic limit is $\epsilon_0 \approx -1.25$, shown here as a dashed line while the critical energy density for the classical ($\Gamma = 0$) glassy transition is $\epsilon_c \approx -1.05$.}
\label{fig:mobility_edge}
\end{center}
\end{figure}

\noindent Using the formula $\Gamma_c(\epsilon) = \exp(-\ln\lvert\psi_N(\epsilon)\rvert /N) = \exp( -Z_N(\epsilon))$ from Eq. \eqref{eq:gamma_c} we obtain a plot of the MBL critical point as a function of the energy density $\epsilon$ shown in Fig. \ref{fig:mobility_edge}. Note that this is  the energy density of the unperturbed eigenstates $\ket{a}$, while usually one would write $\Gamma_c$ as a function of the energy density of the \emph{perturbed} eigenstates $\ket{\psi_\alpha}$. However, the perturbed energies $E_\alpha = E_a + O(\Gamma^2)$ coincide with the unperturbed ones up to second-order corrections in $\Gamma$, which we neglect.

\begin{figure}[htbp]
\begin{center}
\includegraphics[width=0.9\columnwidth]{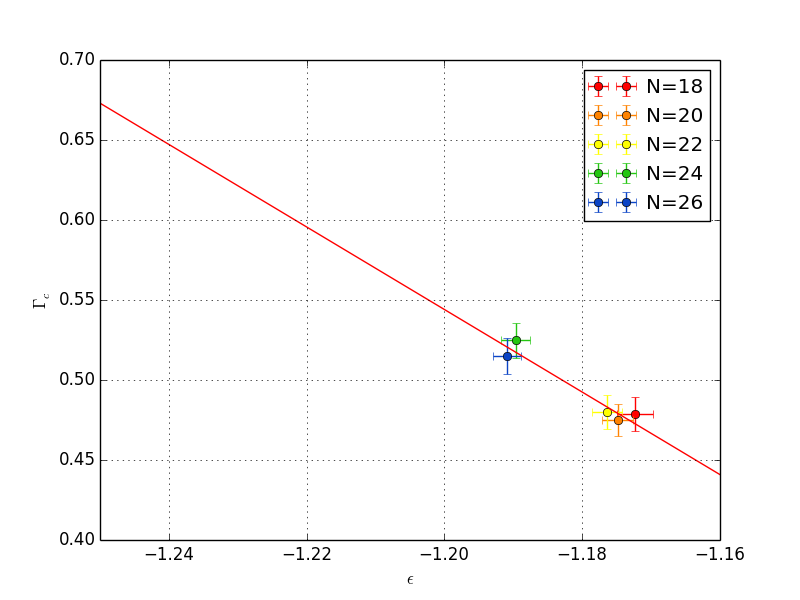}
\caption{Estimates of the $\Gamma_c$ for the ground state ($T=0$ case) for different system sizes $N$. As $N \rightarrow \infty$ the disorder-averaged energy density of the ground state decreases towards the expected thermodynamic limit value of $\epsilon_0 = -1.25$. A linear interpolation of the $\Gamma_c$ values obtained give a thermodynamic limit value of $\Gamma_c = 0.67$.}
\label{fig:ground_state}
\end{center}
\end{figure}

In order to plot the MBL critical line in the $(T,\Gamma)$-phase diagram and compare it to the boundary of the glassy phase, we have to compute the relation $T = T(\epsilon)$ between temperature and (disorder-averaged) energy density. We used standard Monte Carlo methods to extract the (thermal) average energy density of different realizations of disorder at various temperatures and fixed $\Gamma = 0$, then we took the average over the results\footnote{We note that as one approaches the ground state energy, small difference of energies translate to (relatively) large difference of temperatures due to small values of the heat capacity in the low-temperature regime. In order to effectively control this effect one would require better precision in the M.C. energy estimation.}. In order to better understand the low energy regime we studied the ground state of the unperturbed (\ie $\Gamma = 0$) model. For each size $N = 18,20,22,24,26$ we generated a large number ($\geq 1000$) of instances and extracted one of the ground states by performing a thermal annealing (whose results were checked against an exact solver for the smaller sizes). For each ground state we computed the $\Gamma_c$ value using the forward approximation. The disorder-averaged results are show in Fig. \ref{fig:ground_state}. Extrapolations give a value of $\Gamma_c = 0.67$ in the thermodynamic limit, which seems consistent with Fig. \ref{fig:mobility_edge}.

\begin{figure}[htbp]
\begin{center}
\includegraphics[width=0.9\columnwidth]{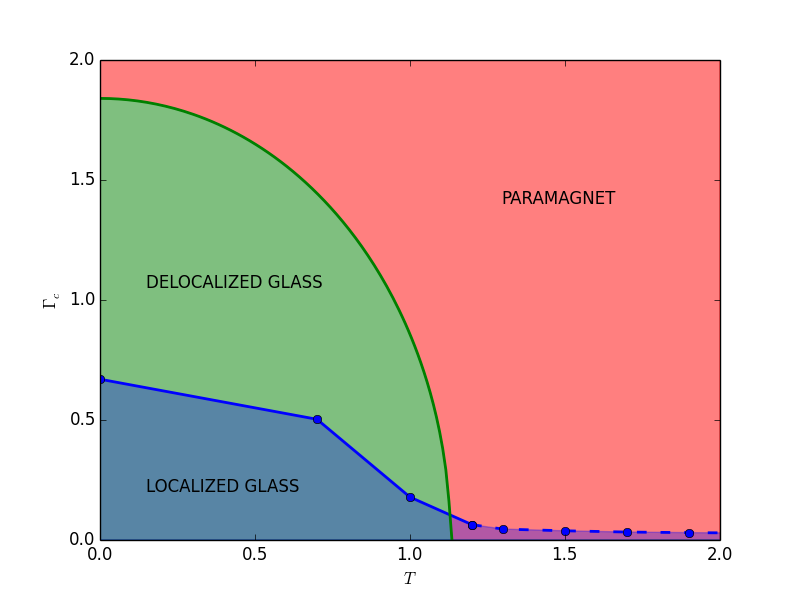}
\caption{Phase diagram of the Hamiltonian of the Ising spin glass in a transverse field (Eq. \eqref{eq:hamiltonian}). The MBL critical line obtained from the numerical data is shown as linked blue dots. The $T=0$ point was obtained from the thermodynamic-limit extrapolation of Fig. \ref{fig:ground_state} while the finite-temperature points were derived from the $\Gamma_c$ values for the largest size ($N=26$) of Fig. \ref{fig:mobility_edge}. With the possible exception of the small, $T > 1.1$ tail the MBL phase seems to be a proper subset of the glassy phase.}
\label{fig:dynamical_phase_diagram}
\end{center}
\end{figure}

\noindent Finally, we plotted a finite-size ($N = 26$) estimate of the MBL critical line and the line of the glassy transition in the $(T,\Gamma)$-phase diagram (Fig. \ref{fig:dynamical_phase_diagram}). The MBL phase seems to be strictly contained in the glassy phase, therefore there is a region of the phase diagram where the system is both glassy and delocalized.

\section{Conclusion}

We studied the localization properties of the transverse-field Ising spin glass model on the $3$-regular random graph in the limit where the trasverse-field is weak compared to the disordered interactions. This model is known to exhibit a transition from a paramagnetic to a glassy phase at low temperatures and weak transverse-field. The classical Ising spin glass model is widely believed to capture the complicated combinatorial structure of general $NP$-hard computational problems while the zero-temperature, weak transverse-field regime describes the final stage of a quantum annealing protocol designed to find the ground-state energy of the Ising spin glass. Many-body localization has been argued to be an obstacle to efficient quantum annealing due to the presence of exponentially-closing gaps in the localized phase.

\noindent We computed numerically the many-body mobility edge of the system in the forward approximation, finding that the energy eigenstates of the system indeed localize for small values of the transverse field at finite system sizes. When plotted against the equibilbrium phase diagram of the model, we discovered that the localized region does not coincide with the glassy phase. In particular, evidence points to the fact that the glassy phase is partitioned into a delocalized region and a localized one. We conjecture that the glassy, delocalized region will exhibit the same clustering of eigenstates observed in \cite{BLPS2} for the $p$-spin model, where the eigenstates were found to form clusters inside of which the energies are distributed according to Wigner-Dyson while the global distribution of the energy levels of the model is Poissonian.

\noindent Moreover, we expect that classical methods that exploit the fine-tuning of thermal relaxation (such as simulated annealing) will perform poorly in the entire glassy phase while quantum annealers will perform poorly only once localization sets in. Therefore we conjecture that in the glassy, delocalized region of the phase space quantum annealing algorithms can outperform any classical thermal annealing protocol.

A natural future direction outlined by our work would be to check whether the same localization/delocalization transition is present when the disordered term of the Hamiltonian encodes a real-life computational problems such as $3$-SAT. In the affirmative case, a detailed comparison of the performance of \eg simulated annealing and quantum annealing (either simulated numerically or by an actual experiment) inside of the region that is both glassy and delocalized would help shed light on the realistic capabilities of quantum annealers over classical thermal annealing and other algorithms based on stochastic local optimization.

\section{Data accessibility}
All numerical data are accessible upon request to the corresponding author.

\section{Competing interests}
The authors declare they have no competing interests.

\section{Authors' contributions}
Both authors conceived and designed the study, G.M.\ carried out the numerical simulations. Both authors contributed in analyzing the data and drafting the manuscript. All authors give final approval for publication.

\section{Acknowledgements}

A.S.\ would like to thank the Google QAI group in Los Angeles (CA, USA), where part of this work was done.

\section{Funding statements}
A.S.\ acknowledges financial support in the form of a Google Faculty Research Award.


\begin{thebibliography}{9}

\bibitem{ColdAtoms}Bloch, I., Dalibard, J. and Zwerger, W., 2008. \textit{Many-body physics with ultracold gases}. Reviews of modern physics, 80(3), p.885.
\bibitem{SCQB} Devoret, M.H. and Martinis, J.M., 2005. \textit{Implementing qubits with superconducting integrated circuits.} In Experimental Aspects of Quantum Computing (pp. 163-203). Springer US.

\bibitem{ETH} Srednicki, M., 1994. \textit{Chaos and quantum thermalization.} Physical Review E, 50(2), p.888. Deutsch, J.M., 1991. \textit{Quantum statistical mechanics in a closed system.} Physical Review A, 43(4), p.2046. D'Alessio, L., Kafri, Y., Polkovnikov, A. and Rigol, M., 2016. \textit{From quantum chaos and eigenstate thermalization to statistical mechanics and thermodynamics.} Advances in Physics, 65(3), pp.239-362.

\bibitem{A58} Anderson, Philip W. "Absence of diffusion in certain random lattices." Physical review 109.5 (1958): 1492.

\bibitem{BAA} Basko, D.M., Aleiner, I.L. and Altshuler, B.L., 2006. \textit{Metal - insulator transition in a weakly interacting many-electron system with localized single-particle states.} Annals of physics, 321(5), pp.1126-1205.

\bibitem{SM} Schiulaz, M., Mueller, M. \textit{Ideal quantum glass transitions: many-body localization without quenched disorder} AIP Conference Proceedings 1610 (1), 11-23 (2014)

\bibitem{DU1} De Roeck, W. and Huveneers, F. (2014). \textit{Scenario for delocalization in translation-invariant systems.} Physical Review B, 90(16), 165137.

\bibitem{PH} Pal, Arijeet, and David A. Huse. "Many-body localization phase transition." Physical review b 82.17 (2010): 174411.

\bibitem{HNO1}Huse, David A., Rahul Nandkishore, and Vadim Oganesyan. "Phenomenology of fully many-body-localized systems." Physical Review B 90.17 (2014): 174202.

\bibitem{ZPP} Znidaric, Marko, Tomaz Prosen, and Peter Prelovsek. "Many-body localization in the Heisenberg X X Z magnet in a random field." Physical Review B 77.6 (2008): 064426.

\bibitem{SPA} Serbyn, Maksym, Z. Papic, and Dmitry A. Abanin. "Local conservation laws and the structure of the many-body localized states." Physical review letters 111.12 (2013): 127201.

\bibitem{I16} Imbrie, John Z. "On many-body localization for quantum spin chains." Journal of Statistical Physics 163.5 (2016): 998-1048.

\bibitem{RMS} Ros, V., M. Mueller, and A. Scardicchio. "Integrals of motion in the many-body localized phase." Nuclear Physics B 891 (2015): 420-465.

\bibitem{IRS} Imbrie, John Z., Valentina Ros, and Antonello Scardicchio. "Review: Local Integrals of Motion in Many-Body Localized systems." arXiv preprint arXiv:1609.08076 (2016).

\bibitem{BDS} Buccheri, Francesco, Andrea De Luca, and Antonello Scardicchio. "Structure of typical states of a disordered Richardson model and many-body localization." Physical Review B 84.9 (2011): 094203.

\bibitem{DS} De Luca, Andrea, and Antonello Scardicchio. "Ergodicity breaking in a model showing many-body localization." EPL (Europhysics Letters) 101.3 (2013): 37003.

\bibitem{LLA} Luitz, David J., Nicolas Laflorencie, and Fabien Alet. "Many-body localization edge in the random-field Heisenberg chain." Physical Review B 91.8 (2015): 081103.

\bibitem{BPM} Bardarson, Jens H., Frank Pollmann, and Joel E. Moore. "Unbounded growth of entanglement in models of many-body localization." Physical review letters 109.1 (2012): 017202.

\bibitem{CKLS} Chandran, Anushya, et al. "Many-body localization and symmetry-protected topological order." Physical Review B 89.14 (2014): 144201.

\bibitem{AKR} Altshuler, Boris, Hari Krovi, and Jeremie Roland. "Anderson localization makes adiabatic quantum optimization fail." Proceedings of the National Academy of Sciences 107.28 (2010): 12446-12450.

\bibitem{FGG} Farhi, Edward, et al. "A quantum adiabatic evolution algorithm applied to random instances of an NP-complete problem." Science 292.5516 (2001): 472-475.

\bibitem{KN} Kadowaki, Tadashi, and Hidetoshi Nishimori. "Quantum annealing in the transverse Ising model." Physical Review E 58.5 (1998): 5355.

\bibitem{KS} Knysh, Sergey, and Vadim Smelyanskiy. "On the relevance of avoided crossings away from quantum critical point to the complexity of quantum adiabatic algorithm." arXiv preprint arXiv:1005.3011 (2010).

\bibitem{LMSS} Laumann, Christopher R., et al. "Quantum annealing: The fastest route to quantum computation?." The European Physical Journal Special Topics 224.1 (2015): 75-88.

\bibitem{LPS} Laumann, C. R., A. Pal, and A. Scardicchio. "Many-body mobility edge in a mean-field quantum spin glass." Physical review letters 113.20 (2014): 200405.

\bibitem{BLPS1} Baldwin, C. L., et al. "The many-body localized phase of the quantum random energy model." Physical Review B 93.2 (2016): 024202.

\bibitem{BLPS2} Baldwin, C. L., et al. "Clustering of non-ergodic eigenstates in quantum spin glasses." arXiv preprint arXiv:1611.02296 (2016).

\bibitem{MPV} Mezard, Marc, Giorgio Parisi, and Miguel Virasoro. Spin glass theory and beyond: An Introduction to the Replica Method and Its Applications. Vol. 9. World Scientific Publishing Co Inc, 1987.

\bibitem{Bethe1935} Bethe, H. A. "Statistical Theory of Superlattices". Proceedings of the Royal Society of London A, 150, 871, pages 552--575, 1935.

\bibitem{Laumann2008} Laumann, C., Scardicchio, A. and Sondhi, S.L. "Cavity method for quantum spin glasses on the Bethe lattice." Phys Rev B, 78, 13, 2008 

\bibitem{Farhi2012} Farhi, E., et al. "Performance of the quantum adiabatic algorithm on random instances of two optimization problems on regular hypergraphs." Phys. Rev. A, 86, 5, 2012.

\bibitem{Mezard2001} M{\'e}zard, M. "The Bethe lattice spin glass revisited." The European Physical Journal B, 20, 2, pages 217--233, 2001.

\bibitem{Mossi2016} Mossi G, et al. "On the quantum spin glass transition on the Bethe lattice." JSTAT, 1, 2017.

\bibitem{VianaBray85} Viana L. and Bray J. "Phase diagrams for dilute spin glasses." Journal of Physics C, 18, 15, 1985.

\bibitem{A-CAT1}Abou-Chacra, R., D. J. Thouless, and P. W. Anderson. "A selfconsistent theory of localization." Journal of Physics C: Solid State Physics 6.10 (1973): 1734.

\bibitem{Aizenman2013} Aizenman M. and Warzel S. "Resonant delocalization for random Schroedinger operators on tree graphs."" Journal of the European Mathematical Society, 15, 4, pages 1167--1222, 2013

\bibitem{Pietracaprina2016} Pietracaprina F., et al. "Forward approximation as a mean-field approximation for the Anderson and many-body localization transitions." Physical Review B, 93, 054201 (2016)

\end{thebibliography}
\end{document}